\documentclass [12pt]{article}
\usepackage[utf8]{inputenc}
\usepackage{authblk}
\usepackage{natbib}
\usepackage{babel}
\usepackage{graphicx}
\usepackage{adjustbox}
\usepackage{tabularx}
\usepackage{geometry}
\usepackage{setspace} 
\usepackage{caption}
\usepackage{textcomp}
\usepackage{subcaption}
\usepackage{hyperref}
\hypersetup{
	colorlinks,
	linkcolor={red!50!black},
	citecolor={blue!50!black},
	urlcolor={blue!80!black}
}

\usepackage{tikz}
\usetikzlibrary{
  arrows.meta,
  patterns,
  decorations.pathreplacing,
  calc
}

\usepackage{newtxtext,newtxmath}

\usepackage{amsmath,amssymb}[noamsbb]
\usepackage{bm}
\usepackage{longtable}
\usepackage{multicol}
\usepackage{multirow}
\usepackage{booktabs}
\usepackage{makecell}
\usepackage{geometry}
\usepackage{comment}
\usepackage{xcolor}
\usepackage{soul}
\usepackage{pgfplots}
\pgfplotsset{compat=1.17} 
\usepackage{array}
\usepackage{float}

\title{Leveraging US Commercial Adjustments:\\ MFN Tariffs and Trade Diversion}
\author[1, 3]{Maria Cipollina\thanks{ \ This paper is conceived as a tangible way to honour the memory of Luca Salvatici, a friend and a coauthor that inspired our research and with his attitude and good humour influenced the way in which we leave our professional and personal live. We thank AIEAA and the organizers of the 15th AIEAA Annual Conference, held in Rome and hosted by Universitas Mercatorum, for the opportunity to plan and realize this contribution.}}
\author[2, 3, 4, 5]{Luca De Benedictis}

\affil[1]{\small University of Molise - cipollina@unimol.it;}
\affil[2]{\small Universitas Mercatorum - luca.debenedictis@unimercatorum.it;}
\affil[3]{Centro Rossi-Doria, Univerisità Roma Tre}
\affil[4]{\small Luiss University}
\affil[5]{\small CRIAVET }
\date{}

\begin{document}

\maketitle
\begin{abstract}
Most-favored-nation (MFN) tariff increases do more than raise trade costs. By changing the relative tariff treatment of competing exporters, they increase the value of existing trade preferences, reshaping relative market access and trade allocation. While the trade-reducing effects of higher tariffs are well established, much less is known about this mechanism. We investigate whether higher MFN tariffs amplify the competitive advantage of preferential partners, generating trade diversion. Using domestic and international trade data, we estimate a structural gravity model with CES-consistent bilateral tariff margins and examine recent changes in US trade policy as an empirical application. We find that higher MFN protection simultaneously reduces imports and redirects trade toward preferential exporters. These effects are highly heterogeneous across sectors and regions, with the largest losses concentrated in intermediate and capital goods, highlighting the role of MFN tariffs in reshaping relative market access and global trade patterns.

\end{abstract}

\section{Introduction}

Most-Favoured-Nation (MFN) tariffs affect international trade through more than a simple increase in trade costs. By changing the relative tariff treatment faced by competing exporters, they also modify the value of existing trade preferences and, consequently, the distribution of market access across suppliers. While higher MFN tariffs are generally expected to reduce imports and encourage domestic production, they may simultaneously strengthen the competitive position of exporters benefiting from preferential trade agreements, inducing a reallocation of trade across countries. Understanding this interaction between protectionism and preferential market access is the purpose of our work, and we believe this topic has become increasingly important today given a trading system characterized by overlapping preferential agreements and renewed trade tensions.

Indeed, the recent increase in US protectionism provides a particularly relevant setting to study these mechanisms. Beginning in 2018, the Trump administration introduced substantial tariff increases through Sections 201, 232, and 301, representing the largest rise in US import protection in decades. Beyond their direct effects on bilateral trade, these measures have become part of a broader process of geoeconomic fragmentation, in which trade policy is increasingly used to pursue strategic and geopolitical objectives rather than purely economic ones. Recent structural gravity evidence shows that fragmentation along geopolitical lines can substantially reduce international trade, alter the geographical distribution of trade flows, and generate heterogeneous welfare losses across countries \citep{campos2023}. At the same time, recent evidence indicates that the consequences of tariff changes extend well beyond trade flows, propagating through production networks, input--output linkages, and firms' sourcing decisions, with important implications for resource allocation and structural transformation (e.g. \citealp{kebede2026}; \citealp{kilumelume2026}). Together, these findings suggest that the effects of protectionism are not limited to higher bilateral trade costs but also reflect broader changes in the competitive environment faced by exporters.

However, while the existing literature documents the aggregate consequences of geopolitical fragmentation and higher tariffs, much less attention has been devoted to the mechanisms through which individual trade policy instruments reshape relative market access. In particular, little is known about how increases in MFN tariffs modify the relative competitive position of exporters by changing the value of existing preferential trade agreements, thereby influencing the reallocation of trade across countries.

This paper addresses a simple but largely unexplored question: \emph{since increases in MFN tariffs do not affect all exporters equally, since they also amplify the competitive advantage of exporters benefiting from preferential agreements, how do trade patterns change?}

Specifically, in this paper, we argue that MFN tariff increases generate a mechanism that we refer to as \emph{preference amplification}. Because preferential tariffs typically remain unchanged while MFN tariffs increase, exporters benefiting from preferential trade agreements experience an improvement in their relative market access even in the absence of additional trade liberalization. As a consequence, imports may be reallocated from non-preferred exporters toward preferential partners through trade diversion, despite an overall contraction in trade. In this perspective, protectionism and preferential trade agreements are not independent policy instruments but interact in shaping bilateral trade flows and market access.

The extent to which this reallocation offsets the contractionary effects of higher protection remains an empirical question. To address it, we estimate a structural gravity model using domestic and international trade data from the USITC International Trade and Production Database for Estimation (ITPDE), combined with MFN and applied tariff data from WITS–TRAINS. Building on the literature on preferential trade policy, trade diversion, and structural gravity models \citep{cipollina2017,cipollina2020,cipollina2025}, we measure exporters' relative market access through CES-consistent bilateral tariff margins. Unlike conventional preference margins, this measure accounts for the tariff treatment faced by all competing suppliers, providing a theoretically consistent representation of multilateral competition.

This paper draws on three main strands of literature. The first examines the effects of recent US protectionist policies and Trump's tariffs on trade, welfare, employment, and third-country outcomes. Recent contributions include \cite{JAVORCIK2025104125} and \cite{CAVALCANTI2026}, who document substantial import reductions, higher costs for US firms and consumers, limited employment gains, and significant spillover effects on third countries. The second strand focuses on trade elasticities, preferential agreements, and preference margins \citep{bouet2005,francois2006,cipollina2017,cipollina2020}. This literature highlights how preferential agreements alter relative market access and generate trade diversion, with recent applications including \citep{cipollina2025,Tamberi2026MFN}. The third strand is the structural gravity literature, which provides the theoretical and empirical framework for evaluating trade policy in the presence of multilateral resistance terms \citep{anderson2003,head2014,yotov2016,fontagne2022, Yotov2022, Yotov2024}.

Integrating these strands, this study demonstrates that the MFN tariff increases operate through a channel that has received little attention in the existing literature. In addition to increasing trade costs, these increases alter the value of existing trade preferences, reshaping relative market access among exporters. This mechanism, which we term \emph{preference amplification} and illustrate in Figure \ref{fig:mfn_protection}, helps explain why protectionism simultaneously generates trade diversion and contraction, redistributing trade opportunities among exporters while simultaneously reducing aggregate trade.

The paper contributes to the literature in three ways. First, it provides new evidence that MFN tariff increases affect not only the volume of trade but also its geographical composition by redistributing relative market access across exporters. Second, it introduces the concept of \emph{preference amplification}, showing that increases in MFN tariffs can strengthen the value of existing trade preferences even when preferential tariffs remain unchanged. Third, it incorporates CES-consistent bilateral tariff margins into a structural gravity framework \citep{anderson2003,yotov2016, Yotov2024}, providing a theoretically grounded measure of relative market access for evaluating the trade effects of protectionist policies.

Our empirical analysis shows that higher MFN protection simultaneously generates trade diversion and trade contraction. Imports are reallocated toward preferential partners, creating winners and losers across sectors, regions, and income groups. However, the gains associated with improved relative market access are insufficient to compensate for the overall decline in trade. The largest losses are concentrated in intermediate and capital goods, suggesting that higher tariffs disrupt production networks and global value chains more than final consumption. These findings indicate that protectionism affects international trade not only by increasing trade costs but also by reshaping the competitive structure of market access.

The remainder of the paper is organized as follows. Section 2 discusses the literature related to our study. Section 3 develops the theoretical framework and introduces the mechanism of \emph{preference amplification}. Section 4 describes the data, while Sections 5 and 6 present the empirical strategy and use the recent US tariff increases as an empirical case study to quantify the effects of the proposed mechanism. Section 7 concludes.

\section{Trade Protectionism and Related Literature}

The recent return to protectionist trade policies has attracted much interest in understanding how tariff increases affect international trade, market access, and global supply chains. Beginning in 2018, the United States introduced a series of tariff measures under Sections 201, 232, and 301 of its trade law, targeting a broad range of imported products and trading partners. Representing the largest increase in US trade protection in decades, these measures have generated an extensive literature examining their effects on trade flows, prices, welfare, employment, and production networks.

Recent research places these developments within the broader context of geoeconomic fragmentation and the reorganization of international production. Rather than simply reducing bilateral trade, higher tariffs have contributed to the diversification of sourcing strategies, the relocation of production across countries, and the restructuring of global value chains. More broadly, the recent resurgence of economic nationalism has reinforced the use of trade policy as a strategic instrument in an increasingly fragmented global economy \citep{obstfeld2021}. In this context, trade policy increasingly shapes the geography of international production by influencing firms' sourcing decisions and the relative competitiveness of exporters, rather than operating solely through changes in bilateral trade costs \citep{freund2024,imf2023fragmentation,wto2023}. At the same time, the coexistence of protectionist measures and preferential trade agreements has created a more fragmented trading environment in which changes in MFN tariffs may generate consequences extending well beyond the countries directly affected.

A first strand of literature examines the economic consequences of recent US protectionist policies and the tariff increases introduced during the Trump administration. Early studies consistently show that these measures reduced imports from targeted countries while imposing substantial costs on US firms and consumers through higher import prices and input costs. Using detailed evidence on tariff incidence, \citet{Amiti} find that the burden of the tariffs was largely borne by US importers rather than foreign exporters, while \citet{fajgelbaum2020} document sizeable welfare losses and show that foreign retaliation amplified the adverse effects of protectionism.

Subsequent research extends the analysis beyond trade and welfare. \citet{JAVORCIK2025104125} find little evidence that the US--China trade war generated meaningful employment gains in the United States, whereas \citet{CAVALCANTI2026} show that trade diversion created employment opportunities in third countries. More recently, \citet{beames2026} highlight substantial sectoral heterogeneity in the effects of recent US tariff measures, reflecting differences in production structures and international supply chains.

An emerging literature further argues that the consequences of higher tariffs extend beyond bilateral trade flows. \citet{freund2024} show that firms responded to the US--China trade war by diversifying suppliers and redirecting imports toward third countries, with only limited reshoring of production. Similarly, \citet{kim2025} argues that recent tariff policies have reshaped global production networks and sourcing strategies, reinforcing the view that protectionism increasingly operates through changes in international production and trade patterns rather than solely through higher bilateral trade costs.

Despite these important contributions, the existing literature focuses primarily on the aggregate effects of tariff increases on trade, welfare, employment, and global value chains. Much less attention has been devoted to how MFN tariff increases modify the relative competitive position of exporters by changing the value of existing preferential trade agreements. This paper addresses this question by examining how higher MFN tariffs reshape relative market access and generate preference amplification.

A second strand of literature examines preferential trade agreements, market access, trade diversion, and trade elasticities. Since the seminal contribution of \citet{viner1950}, economists have recognized that preferential trade policies generate both trade creation and trade diversion by altering the relative competitiveness of exporters. Subsequent theoretical and empirical studies show that preferential agreements reshape bilateral trade flows by changing exporters' relative market access vis-à-vis competing suppliers \citep{bouet2005,francois2006}.

More recently, \citet{cipollina2017,cipollina2020,cipollina2025} argue that conventional measures of preference margins provide only a partial representation of market access because they neglect the tariff treatment faced by competing exporters. To overcome this limitation, they develop CES-consistent bilateral tariff margins that explicitly account for the multilateral tariff structure confronting all suppliers serving the same destination market. This approach highlights that the value of preferential access is inherently relative, depending not only on the tariff concession granted to a particular exporter but also on the tariff treatment applied to competing suppliers. Consistent with this perspective, \citet{vanbiesebroeck2026} show that export growth responds to the effective market-access advantage generated by preferential treatment rather than to the formal existence of preferential agreements alone.

An important implication of this literature is that changes in MFN tariffs may substantially modify the competitive value of existing preferential agreements even when preferential tariff schedules remain unchanged. This insight has recently been applied to the analysis of MFN tariff reforms. Examining the consequences of Brexit, \citet{cipollina2025} show that the introduction of the UK Global Tariff reduced MFN rates for many products, thereby eroding the preference margins previously enjoyed by some exporters and generating significant trade reallocation effects. Similarly, \citet{Tamberi2026MFN} demonstrate that changes in MFN tariffs substantially affect import patterns and estimate import elasticities considerably larger than those commonly reported in the literature. Together, these studies show that modifications to MFN tariffs alter the value of existing trade preferences even when preferential agreements themselves remain unchanged.

A closely related branch of this literature concerns the estimation of trade elasticities, which determine the responsiveness of trade flows to changes in trade costs and constitute a key ingredient of quantitative trade models and counterfactual analyses. Using worldwide bilateral trade data, \citet{fontagne2022} estimate product-level trade elasticities and document substantial heterogeneity across products and sectors. More recently, \citet{Tamberi2026MFN} show that explicitly accounting for MFN tariff exposure and preference utilization leads to considerably larger import elasticities than those commonly reported in the literature. These findings suggest that the response to tariff changes varies substantially across products and that accurate elasticity estimates are essential for evaluating trade policy reforms.

This heterogeneity is particularly relevant when MFN tariffs increase. Differences in substitution elasticities, production structures, and participation in global value chains imply that identical tariff increases may generate markedly different responses across sectors. Consequently, the magnitude of both trade diversion and trade contraction depends critically on the interaction between changes in relative market access and the substitutability of competing foreign suppliers.

These contributions have substantially improved our understanding of how preferential agreements shape market access, how MFN tariff reforms affect the value of existing preferences, and how trade elasticities govern the response of imports to changes in trade costs. Nevertheless, the literature has focused predominantly on the effects of tariff liberalization or MFN tariff reductions. Much less attention has been devoted to the opposite policy experiment, namely how increases in MFN tariffs modify the effective value of existing trade preferences by reshaping the relative competitiveness of exporters.

This paper addresses this question. Whereas previous studies examine how reductions in MFN tariffs erode existing preference margins, we investigate how increases in MFN tariffs strengthen them. By raising the tariff disadvantage faced by non-preferred exporters while leaving preferential tariffs unchanged, higher MFN protection increases the relative market access of preferential partners and may generate trade diversion without the introduction of new preferential agreements. We refer to this mechanism as preference amplification, namely the increase in the value of existing trade preferences resulting from higher MFN protection rather than from additional preferential liberalization.

A third strand of literature concerns the structural gravity model, which has become the benchmark framework for analysing the effects of trade policy on bilateral trade flows. A key insight of the structural gravity model is that bilateral trade depends not only on direct bilateral trade costs but also on the trade barriers faced by all competing exporters serving the same destination market. As formalized by \citet{anderson2003}, multilateral resistance terms are essential for obtaining theory-consistent estimates of trade policy effects because they capture the relative rather than absolute nature of market access. This insight has fundamentally reshaped empirical trade analysis and motivated the widespread use of high-dimensional fixed effects in modern gravity estimations.

Building on this theoretical framework, subsequent contributions have consolidated structural gravity as the standard tool for empirical policy evaluation. \citet{baldwin2006} emphasize the importance of embedding gravity estimations within general equilibrium frameworks when assessing trade policy reforms, while \citet{head2014} provide a comprehensive synthesis of the theoretical foundations and empirical implementation of gravity models. Similarly, \citet{yotov2016} develop a unified framework for estimating structural gravity models and conducting theoretically consistent counterfactual analyses, providing practical guidelines that have become the benchmark for applied trade policy research.

More recent methodological contributions further extend the scope of structural gravity. \citet{Yotov2022} show that incorporating domestic trade flows is essential for theory-consistent estimation and for accurately measuring changes in trade costs and welfare. More broadly, \citet{Yotov2024} review the evolution of structural gravity from an empirical regularity to the workhorse model for evaluating trade policies and conducting counterfactual analysis. Extending these developments, \citet{larch2023} propose a flexible structural gravity framework capable of quantifying both observable and unobservable changes in trade costs in ex-ante policy simulations, highlighting its versatility for evaluating both trade liberalization and protection scenarios.

The present paper builds on this literature by combining the structural gravity framework with the literature on preferential market access. Because structural gravity models trade as a function of relative market access, changes in MFN tariffs affect trade not only by increasing bilateral trade costs but also by modifying the competitive position of all exporters serving the same destination market. Consequently, changes in MFN tariffs alter the effective value of existing trade preferences even when preferential tariff schedules remain unchanged. This multilateral perspective provides the theoretical foundation for the mechanism of preference amplification developed in the next section.

\section{Theoretical Framework}
\label{sec:theory}

The analysis adopts a structural gravity framework with CES preferences in which US consumers allocate expenditure across domestic and imported varieties of each product. Products are differentiated by country of origin, implying that exporters compete through relative prices and trade costs.

Let ($M_{ikt}$) denote US imports of product (k) from exporter (i) in year (t). Under CES preferences, bilateral import demand can be expressed as:

\begin{equation}
M_{ikt} = \alpha_{ikt}P_{ikt}^{-\sigma_k}\Pi_{kt}^{\sigma_k-1}E_{kt},
\end{equation}

where ($E_{kt}$) denotes total expenditure on product (k), ($\sigma_k$) is the elasticity of substitution across exporters, ($P_{ikt}$) is the consumer price of imports from exporter (i), and ($\Pi_{kt}$) is the CES price index across all suppliers.

Consumer prices incorporate tariffs and other trade costs:

\begin{equation}
P_{ikt}=p_{ikt} c_{ikt}(1+\tau_{ikt}),
\end{equation}

where ($p_{ikt}$) is the FOB export price, ($c_{ikt}$) denotes non-tariff trade costs, and ($\tau_{ikt}$) is the tariff applied by the US.

Following \citet{cipollina2017}, further developed in \citet{cipollina2020} and \citet{cipollina2025}, relative market access is measured through bilateral tariff margins:

\begin{equation}
btm_{ikt}=\frac{1+T_{kt}}{1+\tau_{ikt}},
\end{equation}

where $(T_{kt}$) is a CES-consistent reference tariff factor summarizing the tariff structure faced by all exporters supplying product (k) to the US market.

The bilateral tariff margin measures the relative competitiveness of each exporter. Values above one indicate that an exporter faces tariffs lower than the multilateral benchmark and therefore enjoys a competitive advantage. Conversely, values below one indicate a competitive disadvantage relative to competing suppliers.

Unlike conventional preference margins, bilateral tariff margins explicitly account for multilateral trade resistance. Market access therefore depends not only on the tariff applied to a specific exporter, but also on the tariff structure faced by all competitors.

\begin{figure}[H]
\centering
\includegraphics[width=0.95\textwidth]{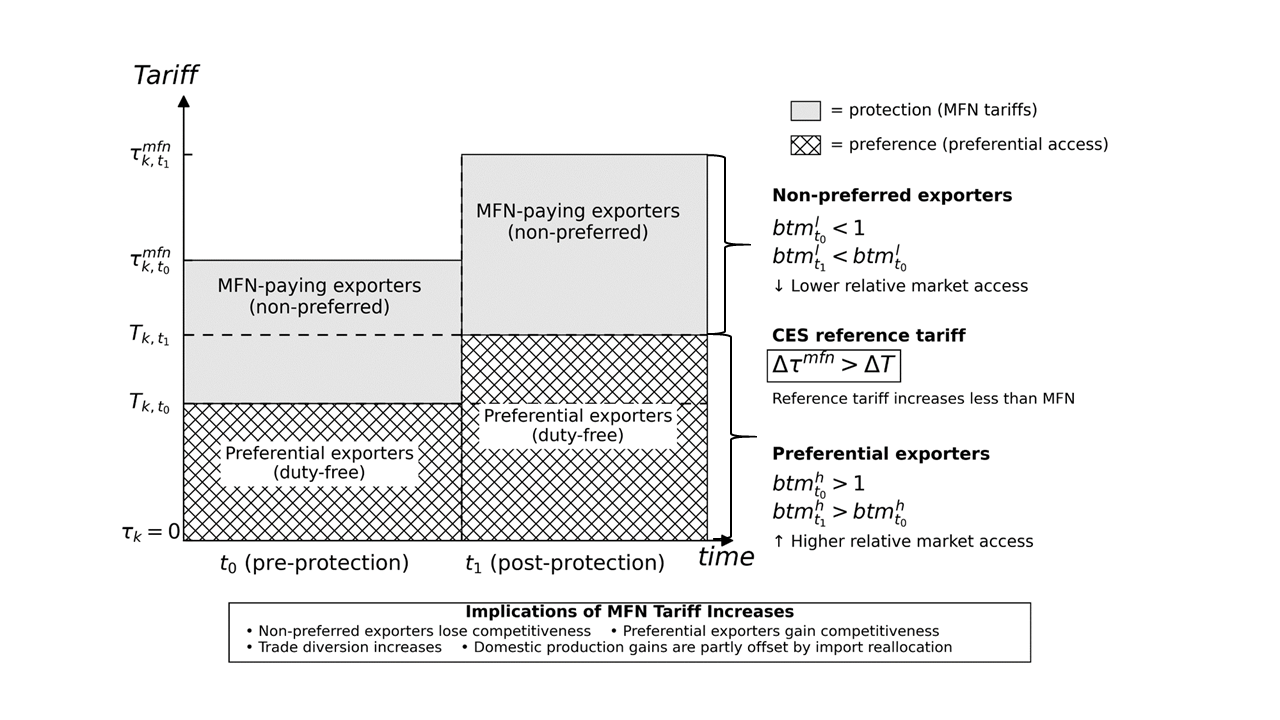}
\caption{MFN protection and the structure of tariff margins: the US case}
\label{fig:mfn_protection}
\end{figure}

Figure \ref{fig:mfn_protection} illustrates the mechanism through which MFN protection affects tariff margins.

At time ($t_0$), before the tariff increase, non-preferred exporters pay the MFN tariff and face tariff margins below one:

\begin{equation}
btm^{l}_{t_0}<1,
\end{equation}

while exporters benefiting from preferential access face lower tariffs and therefore enjoy tariff margins above one:

\begin{equation}
btm^{h}_{t_0}>1.
\end{equation}

Suppose that at time ($t_1$) the United States increases MFN tariffs on non-preferred exporters. The reference tariff factor also increases because it reflects the overall tariff structure faced by exporters. However, since the reference tariff is a CES aggregate, it increases by less than the MFN tariff itself:

\begin{equation}
\tau^{mfn}_{k,t_1}-\tau^{mfn}_{k,t_0}>T_{k,t_1}-T_{k,t_0}.
\end{equation}

As a consequence, tariff margins move further away from unity.

For non-preferred exporters, tariff margins fall:

\begin{equation}
btm^{l}_{t_1}<btm^{l}_{t_0}<1,
\end{equation}

implying a deterioration in relative market access.

For preferential exporters, tariff margins increase:

\begin{equation}
btm^{h}_{t_1}>btm^{h}_{t_0}>1,
\end{equation}

strengthening their relative competitiveness.

This mechanism is illustrated in Figure \ref{fig:mfn_protection}. The left panel represents the pre-protection equilibrium (($t_0$)), while the right panel shows the situation after the MFN tariff increase (($t_1$)). The lower trajectory represents non-preferred exporters (($btm^{l}$)), whereas the upper trajectory represents exporters benefiting from preferential access (($btm^{h}$)).

Following the increase in MFN tariffs, the distance between the two groups widens. Low tariff margins become lower and high tariff margins become higher. Therefore, MFN protection amplifies existing preferences even when preferential tariffs remain unchanged. We refer to this mechanism as \emph{preference amplification}.

An important implication is that MFN protection does not necessarily translate into a proportional increase in domestic production. Part of the adjustment occurs through trade diversion, as imports are reallocated from non-preferred exporters toward countries that retain relatively favourable market access conditions.

Consequently, MFN tariff increases generate two simultaneous effects. First, they reduce imports from non-preferred exporters by worsening their relative competitiveness. Second, they increase the attractiveness of preferential suppliers and redirect trade toward exporters facing more favourable tariff margins.

The magnitude of these effects depends on three factors: the variation in bilateral tariff margins, the elasticity of substitution across exporters, and the initial distribution of trade flows. In sectors characterized by high substitution elasticities, even relatively small changes in tariff margins can generate substantial trade diversion, whereas sectors with limited substitution possibilities are expected to experience larger trade-contraction effects \citep{cipollina2025}.

\section{Data}

The empirical analysis combines information from two complementary sources. Bilateral trade flows and domestic production are obtained from the United States International Trade Commission (USITC) International Trade and Production Database for Estimation (ITPDE), while MFN and preferential applied tariffs are collected from the World Integrated Trade Solution (WITS–TRAINS) database. Tariff information, originally available at the Harmonized System (HS) product level, is matched to the ITPDE industry classification using the HS–ISIC Rev. 3 concordance. The resulting dataset provides consistent information on bilateral imports, domestic production, and tariff protection at the exporter–industry–year level.

The sample covers the period 2016--2021, encompassing the years immediately before and after the major increases in US import protection. The final database combines domestic production with bilateral import flows for all US trading partners, allowing the estimation of a structural gravity model that incorporates both international and domestic trade. For descriptive statistics and counterfactual simulations, industries are further aggregated into broader HS sections and end-use categories.

The dependent variable is US imports by exporter, product, and year. The key explanatory variable is the bilateral tariff margin defined in Section \ref{sec:theory}, complemented by information on MFN and applied tariffs. The empirical specification exploits variation across exporters, products, and time to identify the effects of changes in trade protection on bilateral imports.

Particular attention is devoted to the distinction between exporters benefiting from preferential access and exporters facing MFN tariffs. Since bilateral tariff margins depend on the entire tariff structure rather than on bilateral tariffs alone, changes in MFN protection can modify the relative competitiveness of exporters even when preferential tariffs remain unchanged. Consequently, MFN tariff increases may alter the allocation of imports across suppliers through changes in relative market access.

\begin{figure}[H]
\centering
\includegraphics[width=0.95\textwidth]{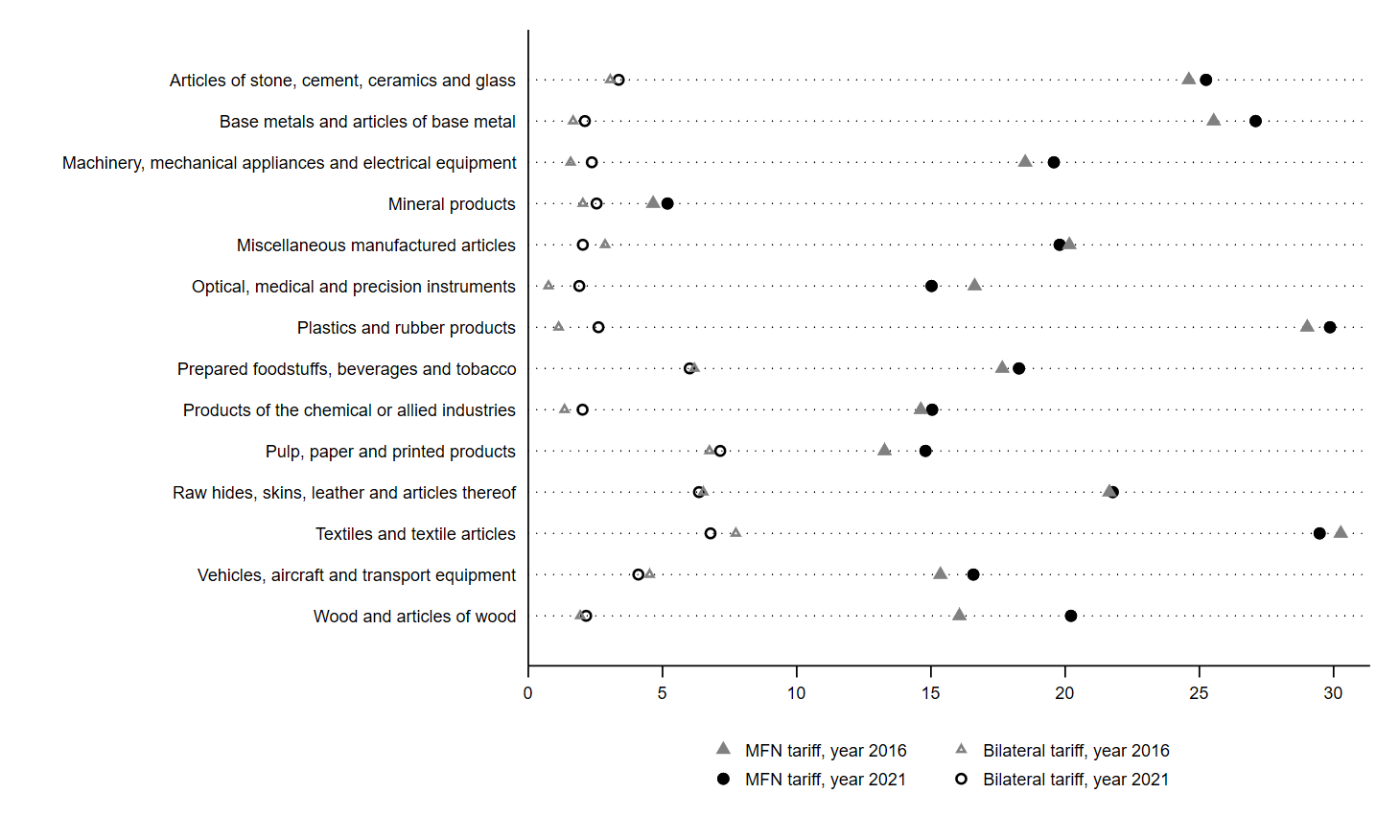}
\caption{Average US MFN tariffs by sector, 2016 and 2021.}
\label{fig:mfn_tariff}
\end{figure}

Figure \ref{fig:mfn_tariff} reports the evolution of average US MFN tariffs between 2016 and 2021. Two patterns clearly emerge. First, MFN tariffs increased substantially in several manufacturing sectors following the tariff measures introduced by the Trump administration, although the magnitude of the increase varies considerably across industries. The largest increases are observed for textiles, plastics and rubber products, base metals, articles of stone, cement, ceramics and glass, and machinery and electrical equipment, while other sectors experienced more moderate changes.

Second, bilateral applied tariffs remained largely unchanged over the same period and continued to be substantially lower than MFN tariffs in most sectors, reflecting the extensive network of preferential trade agreements maintained by the United States. Consequently, the gap between MFN and bilateral tariffs widened considerably between 2016 and 2021.

Using the conventional definition of preference margins---namely the difference between MFN and preferential applied tariffs---Figure \ref{fig:mfn_tariff} suggests that the US already granted sizeable tariff preferences across most product groups in 2016. The subsequent increase in MFN tariffs mechanically enlarged these preference margins, even though preferential tariff schedules remained essentially unchanged. From a descriptive perspective, this implies that the competitive advantage enjoyed by preferential exporters increased without the negotiation of new preferential agreements or additional tariff concessions.

However, the figure also highlights the limitations of interpreting preference margins solely as the difference between MFN and preferential tariffs. Such a measure provides only a partial representation of market access because it ignores the tariff treatment faced by all competing exporters serving the same destination market. As emphasized by the literature on CES-consistent bilateral tariff margins \citep{cipollina2017,cipollina2020,cipollina2025}, the effective value of preferential access depends on the entire tariff structure confronting exporters rather than on a simple bilateral tariff differential.

The empirical analysis developed in this paper therefore moves beyond this descriptive evidence by adopting a theoretically consistent measure of relative market access. Within this framework, increases in MFN tariffs do not merely widen conventional preference margins; they increase the relative competitiveness of preferential suppliers vis-\`a-vis all competing exporters. This mechanism provides the central channel through which higher MFN protection reshapes trade patterns and generates trade diversion toward preferential partners.

\begin{figure}[H]
\centering
\includegraphics[width=0.95\textwidth]{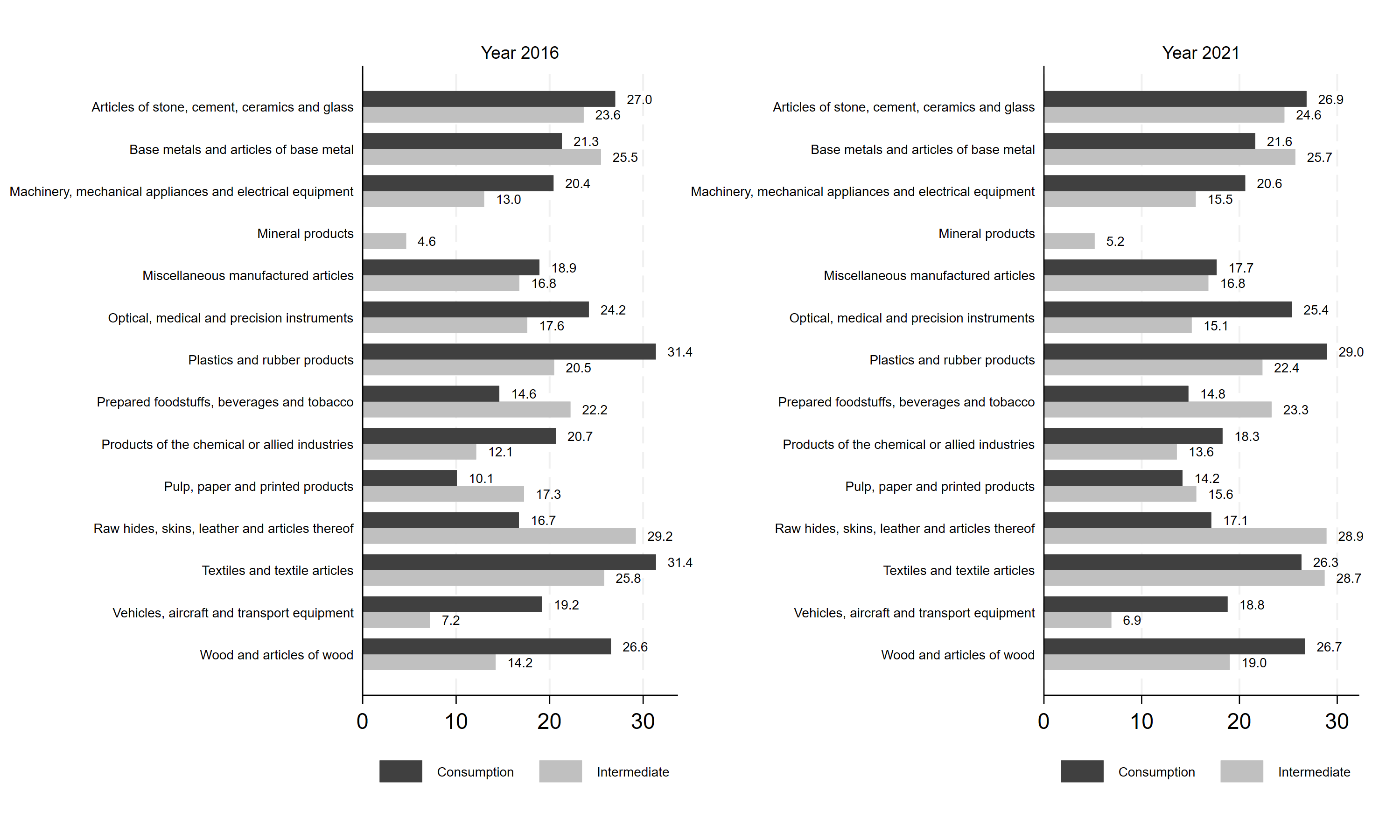}
\caption{Average US MFN tariffs by BEC category, 2016 and 2021.}
\label{fig:tariff_bec}
\end{figure}

To further illustrate the sectoral distribution of protection, Figure \ref{fig:tariff_bec} reports average MFN tariffs by Broad Economic Categories (BEC), distinguishing between consumption and intermediate goods across product groups in 2016 and 2021. The figure reveals substantial heterogeneity both across sectors and across end-use categories. In most industries, consumption goods are subject to higher MFN tariffs than intermediate goods, reflecting the greater protection traditionally granted to final consumer products. Particularly high tariff levels are observed for textiles, plastics and rubber products, articles of stone, ceramics and glass, and wood products, whereas mineral products, pulp and paper, and food-related industries display considerably lower levels of protection.

Although average tariff levels changed only moderately between 2016 and 2021, the cross-sector differences remain pronounced. These differences imply that the increase in US protection did not affect all industries uniformly. Sectors characterized by higher tariff protection experienced a larger increase in the tariff differential between preferential and non-preferential suppliers, thereby creating stronger incentives for trade reallocation toward preferential exporters. Conversely, sectors with relatively low MFN protection are expected to exhibit more limited adjustments in sourcing patterns.

The distinction between consumption and intermediate goods is particularly relevant in the context of recent protectionist policies. Tariffs on intermediate goods affect firms through higher production costs and changes in input sourcing, with potential consequences for production networks and global value chains. By contrast, tariffs on consumption goods primarily influence the relative competitiveness of foreign suppliers in final goods markets. Consequently, identical changes in MFN tariffs may generate markedly different effects across industries depending on the composition of imports, production structures, and the role of imported intermediates.

The figure therefore provides additional motivation for the empirical analysis. The considerable heterogeneity in tariff protection across sectors and end-use categories suggests that the effects of higher MFN tariffs are unlikely to be uniform. Instead, the magnitude of trade diversion and preference amplification is expected to vary systematically across industries according to both the level of protection and the structure of import demand.

The importance of preferential access becomes clearer when considering the composition of US trade across tariff regimes. Table \ref{tab:trade_composition} decomposes international trade into duty-free trade, preferential trade, and MFN-regime trade for each HS section.

\begin{table}[H]
\centering
\caption{Trade composition by tariff regime and HS section}
\label{tab:trade_composition}

\footnotesize

\begin{tabularx}{\textwidth}{lXccccc}
\toprule
HS &
Sector &
Domestic &
\multicolumn{3}{c}{International trade} &
Total trade \\
\cmidrule(lr){4-6}

&
&
trade &
Duty-free &
Preferential &
MFN &
(USD mn) \\
\midrule

 IV & Prepared foodstuffs, beverages and tobacco
& 90\% & 6\% & 75\% & 19\% & 980 \\

 IX & Wood and articles of wood
& 92\% & 32\% & 65\% & 3\% & 118 \\

 V & Mineral products
& 64\% & 52\% & 35\% & 13\% & 472 \\

 VI & Products of the chemical or allied industries
& 68\% & 4\% & 94\% & 2\% & 662 \\

 VII & Plastics and rubber products
& 74\% & 0\% & 100\% & 0\% & 269 \\

 VIII & Raw hides, skins, leather and articles thereof
& 4\% & 0\% & 100\% & 0\% & 40 \\

 X & Pulp, paper and printed products
& 91\% & 21\% & 79\% & 0\% & 268 \\

 XI & Textiles and textile articles
& 25\% & 0\% & 96\% & 4\% & 157 \\

 XIII & Articles of stone, cement, ceramics and glass
& 91\% & 9\% & 91\% & 0\% & 309 \\

 XV & Base metals and articles of base metal
& 70\% & 1\% & 99\% & 0\% & 504 \\

 XVI & Machinery, mechanical appliances and electrical equipment
& 41\% & 0\% & 98\% & 2\% & 790 \\

 XVII & Vehicles, aircraft and transport equipment
& 61\% & 0\% & 86\% & 14\% & 885 \\

 XVIII & Optical, medical and precision instruments
& 56\% & 2\% & 98\% & 0\% & 229 \\

 XX & Miscellaneous manufactured articles
& 49\% & 0\% & 99\% & 0\% & 235 \\

\bottomrule
\end{tabularx}

\vspace{0.2cm}

\footnotesize
\textit{Notes:} Domestic trade is expressed as a share of total trade. The international trade columns report the distribution of international trade across duty-free, preferential, and MFN regimes. Total trade is measured in millions of US dollars.
\end{table}

Table \ref{tab:trade_composition} reveals that preferential trade is the dominant regime across most manufacturing sectors. In chemicals, textiles, machinery, optical instruments, base metals, and miscellaneous manufactures, between 94 and 99 percent of international trade takes place under preferential tariff arrangements. In particular, preferential trade accounts for 98 percent of imports in machinery and optical products, 99 percent in base metals and miscellaneous manufactures, and 96 percent in textiles.

By contrast, imports subject to MFN tariffs account for a relatively small share of trade in most industries and remain quantitatively important only in a limited number of sectors. The highest MFN shares are observed in prepared foodstuffs (19 percent), vehicles and transport equipment (14 percent), and mineral products (13 percent). In most other manufacturing sectors, MFN trade represents less than 5 percent of international imports.

Duty-free trade is concentrated in a few sectors, particularly mineral products, where it accounts for 52 percent of international trade, followed by wood products (32 percent) and pulp and paper products (21 percent). In contrast, duty-free trade is virtually absent in most manufacturing industries, where preferential arrangements dominate market access.

The table also highlights substantial differences in the relative importance of domestic production. Domestic trade exceeds 90 percent of total trade in prepared foodstuffs, wood products, and stone, glass and ceramic products, whereas it represents only 25 percent in textiles and 41 percent in machinery. These differences suggest varying degrees of international integration across sectors and may influence the extent to which tariff changes translate into trade diversion or substitution toward domestic production.

The evidence indicates that most US imports already benefit from preferential treatment. Consequently, increases in MFN tariffs primarily affect relative market access across exporters rather than the tariff treatment of the majority of existing trade flows. Changes in bilateral tariff margins therefore constitute an important channel through which protection can reshape trade patterns and reallocate imports across trading partners.

\section{Empirical Strategy}

To quantify the effects of changes in relative market access on US imports, we estimate product-level substitution elasticities within a structural gravity framework using the Poisson Pseudo-Maximum Likelihood (PPML) estimator. PPML provides consistent estimates in the presence of heteroskedasticity and naturally accommodates zero trade flows.

The empirical specification relates bilateral imports to the bilateral tariff margin introduced in Section \ref{sec:theory}. For each product (k), the estimating equation is:

\begin{equation}
E[M_{ikt}|X]=\exp\left[\beta_k \ln(btm_{ikt})+\gamma_{it}+\delta_{kt}
\right],
\end{equation}

where ($M_{ikt}$) denotes US imports of product (k) from exporter (i) in year (t), ($btm_{ikt}$) is the bilateral tariff margin, and ($\beta_k$) is the product-specific elasticity of substitution. The coefficient ($\beta_k$) measures the responsiveness of imports to changes in relative market access generated by tariffs.

The specification includes exporter--year fixed effects (($\gamma_{it}$)) and product--year fixed effects (($\delta_{kt}$)). Exporter--year fixed effects absorb all time-varying exporter-specific determinants of trade, including productivity, exchange rates, macroeconomic conditions, and country-specific policy changes. Product--year fixed effects control for product-specific demand shocks, common policy developments, and changes in US market conditions affecting a given product.

The analysis is conducted from the perspective of a single importing country, namely the United States. Consequently, the specification cannot include the full set of importer--time or importer--product--time fixed effects typically employed in multi-country structural gravity models to control for all general-equilibrium effects. The estimated elasticities should therefore be interpreted within a partial-equilibrium framework. Nevertheless, the combination of exporter--year and product--year fixed effects captures a large share of the relevant variation in exporter-specific and product-specific conditions affecting bilateral trade flows.

Standard errors are clustered at the exporter--product level to account for serial correlation within bilateral trade relationships over time. Identification relies on differences in bilateral tariff margins across exporters within the same product-year. Conditional on the fixed effects, the estimated elasticity captures the effect of changes in relative market access on bilateral imports.

The estimation proceeds iteratively in order to ensure consistency between the estimated substitution elasticities and the tariff margins used in the gravity equation.

In the first step, bilateral tariff margins are computed using MFN tariffs as the reference tariff factor. The resulting elasticity estimates are then used to construct CES-consistent reference tariff factors:

\begin{equation}
T_{kt}=\left(\sum_i\alpha_i(1+\tau_{ikt})^{1-\sigma_k}\right)^{\frac{1}{1-\sigma_k}},
\end{equation}

where ($\sigma_k$) denotes the elasticity of substitution for product (k), ($\tau_{ikt}$) is the tariff applied to exporter (i), and ($\alpha_i$) represents the CES distribution parameter.

The CES reference tariff factor summarizes the multilateral tariff structure faced by all exporters supplying product (k) to the US market. In this way, the bilateral tariff margin incorporates an explicit measure of relative market access that accounts for the tariff treatment of competing suppliers rather than relying solely on bilateral tariff preferences.

Using the estimated elasticities, updated CES reference tariff factors are computed and new bilateral tariff margins are obtained. The gravity equation is then re-estimated and the procedure is repeated until convergence. The iterative procedure ensures consistency between substitution elasticities and tariff margins.

While the CES tariff margins incorporate an important component of multilateral trade resistance through the tariff structure faced by competing exporters, they do not fully capture all general-equilibrium effects that would be observable in a multi-country framework. Accordingly, the resulting estimates should be interpreted as partial-equilibrium measures of the trade effects generated by changes in relative market access.

The final tariff margins are therefore theoretically consistent with the estimated substitution elasticities and capture the multilateral tariff environment faced by exporters in the US market. Unlike conventional preference margins, the resulting measure reflects the relative competitiveness of exporters within the entire tariff structure rather than within a purely bilateral framework.

\subsection{Counterfactual Analysis}

The estimated elasticities are subsequently used to evaluate the trade effects of the tariff measures announced by the United States in February 2026.

For each bilateral trade flow, the simulated tariff is obtained by adding the relevant surcharge to the observed tariff:

\begin{equation}
\tau^{sim}_{ikt}=\tau_{ikt}+rate_{ikt},
\end{equation}

where ($rate_{ikt}$) denotes the tariff increase associated with the policy scenario.

Using the simulated tariffs and the estimated substitution elasticities, CES reference tariff factors and bilateral tariff margins are recomputed under the counterfactual tariff structure. These updated margins are then introduced into the gravity model to generate a new set of predicted imports, denoted by ($TradePrefSIM_{ikt}$).

Following \citet{lai2004}, the proportional trade effect of the tariff reform is measured as:

\begin{equation}
\Delta Trade_{ikt} = \frac{TradePrefSIM_{ikt}-TradePref_{ikt}}{TradePref_{ikt}},
\end{equation}

where ($TradePref_{ikt}$) denotes predicted imports under the observed tariff structure.

The corresponding effect in value terms is obtained by applying the estimated proportional variation to observed trade flows:

\begin{equation}
TradeEffect_{ikt} =Trade_{ikt}\left(\frac{TradePrefSIM_{ikt}-TradePref_{ikt}}{TradePref_{ikt}}\right).
\end{equation}

The counterfactual exercise compares trade outcomes under the observed tariff structure with those implied by the February 2026 tariff schedule. The framework therefore isolates the effects of changes in MFN protection on bilateral tariff margins and, consequently, on relative market access across exporters.

Because tariff margins are defined relative to the multilateral tariff structure, the simulations capture both trade contraction and trade diversion effects. Higher MFN tariffs may reduce aggregate imports while simultaneously reallocating trade toward exporters that retain relatively favorable market access. Aggregating bilateral effects across products, sectors, countries, regions, and income groups provides a comprehensive assessment of the distributional consequences of the tariff reform.

\section{Results}

Table \ref{tab:sigma_sector} reports the estimated substitution elasticities by HS section. The elasticities are first estimated at the HS6 product level using PPML regressions based on CES bilateral tariff margins and are subsequently aggregated at the sector level using simple averages. The first column reports averages computed across all product-level estimates, while the second column restricts attention to coefficients that are statistically significant at the 10 percent level.

\begin{table}[H] 
\centering 
\small 
\caption{Estimated elasticities of substitution by HS section} \label{tab:sigma_sector} 
\begin{tabular}{llcc} 
\hline HS & Sector & All  & Significant  \\ 
 &  & estimates &  estimates \\ 
\hline IV & Prepared foodstuffs, beverages and tobacco & 27.910 & 41.901 \\ 
V & Mineral products & 6.320 & 7.424 \\ 
VI & Products of the chemical or allied industries & 33.908 & 57.864 \\ VII & Plastics and rubber products & 46.638 & 46.638 \\ 
VIII & Raw hides, skins, leather and articles thereof & 4.481 & 5.627 \\ IX & Wood and articles of wood & 5.007 & 4.085 \\ 
X & Pulp, paper and printed products & 2.000 & 2.000 \\ 
XI & Textiles and textile articles & 6.961 & 9.354 \\ 
XIII & Articles of stone, cement, ceramics and glass & 6.839 & 7.233 \\ XV & Base metals and articles of base metal & 17.334 & 26.787 \\ 
XVI & Machinery, mechanical appliances and electrical equipment & 9.704 & 11.650 \\ 
XVII & Vehicles, aircraft and transport equipment & 13.629 & 16.820 \\ XVIII & Optical, medical and precision instruments & 64.327 & 64.327 \\ XX & Miscellaneous manufactured articles & 8.538 & 10.032 \\
\hline \end{tabular} \vspace{0.2cm}
\begin{minipage}{\textwidth}
\footnotesize
\textit{Notes:} The table reports simple averages of product-level substitution elasticities estimated separately for each HS6 product using PPML. Negative estimates are replaced by a value of 2, consistent with the CES assumption that substitution elasticities must exceed unity. The second column reports averages computed using only coefficients statistically significant at the 10\% level.
\end{minipage}

\end{table}

The results reveal substantial heterogeneity in substitution patterns across sectors. Average elasticities computed using all product-level estimates range from 2.0 in pulp, paper and printed products to more than 64 in optical, medical and precision instruments. High elasticities are also observed in plastics and rubber products (46.6), chemicals (33.9), and prepared foodstuffs (27.9).

These differences suggest that the responsiveness of imports to changes in relative market access varies considerably across sectors. In industries characterized by high elasticities, import demand can be more easily reallocated across suppliers following changes in bilateral tariff margins. Consequently, tariff-induced changes in competitiveness are expected to generate stronger trade diversion effects. By contrast, sectors such as wood products, mineral products, leather products, and paper products exhibit relatively low elasticities, indicating more limited substitution possibilities among exporters.

Restricting attention to statistically significant estimates generally increases the average elasticity within sectors. The largest differences are observed for chemicals, prepared foodstuffs, and base metals, where the average elasticity rises substantially when insignificant coefficients are excluded. This pattern suggests that products exhibiting stronger responses to changes in tariff margins are also those for which the relationship between trade flows and relative market access is estimated more precisely.

The magnitude and dispersion of the estimated elasticities are broadly consistent with the empirical trade literature, which documents substantial heterogeneity in substitution patterns across products and sectors. Using a global dataset of bilateral trade flows and tariffs, \citet{fontagne2022} estimate product-level trade elasticities and find considerable variation across HS6 products, with relatively low elasticities for some differentiated goods and substantially larger values for more homogeneous products. While our estimates are not directly comparable because they are obtained from a single-importer framework and rely on CES-consistent tariff margins, they display a similar degree of cross-sector heterogeneity.

The estimated elasticities play a central role in the construction of CES-consistent tariff margins and in the counterfactual simulations presented below. Sectors characterized by higher elasticities are expected to exhibit larger trade reallocation effects following changes in MFN protection, whereas sectors with lower elasticities are likely to experience more limited trade diversion and a greater proportion of trade contraction.

\subsection{Descriptive evidence on CES bilateral tariff margins}
Table \ref{tab:sector_btm} reports average CES bilateral tariff margins by HS section in 2016 and 2021. Overall, the average margin declined slightly from 0.990 to 0.988, indicating a modest deterioration in relative market access between the two years.

Considerable heterogeneity emerges across sectors. The largest declines are observed in mineral products, where the average tariff margin falls from 0.993 to 0.982, and in machinery and electrical equipment, where it decreases from 0.997 to 0.990. Smaller reductions are also observed in chemicals, plastics and rubber products, wood products, and articles of stone, cement, ceramics and glass. These patterns suggest that changes in the tariff structure affected sectors differently, modifying the relative competitiveness of exporters across industries.

A few sectors exhibit stable or slightly improving tariff margins. Textiles increase from 0.992 to 0.998, miscellaneous manufactured articles from 0.987 to 0.995, and vehicles and transport equipment from 0.972 to 0.974. Although the changes are relatively small, they indicate that some sectors experienced a more favourable evolution in relative market access than others.

Average tariff margins exceed unity only in a limited number of sectors. Raw hides and leather products display the highest margins, increasing from 1.026 to 1.028, while optical, medical and precision instruments remain above one despite a decline from 1.022 to 1.012. In these sectors, exporters benefit from tariffs that are, on average, lower than the CES reference tariff, implying a relative competitive advantage compared with competing suppliers.

\begin{table}[H]
\centering
\small
\caption{Average CES bilateral tariff margins by HS section}
\label{tab:sector_btm}

\begin{tabular}{llcc}
\hline
HS & Sector & 2016 & 2021 \\
\hline

IV & Prepared foodstuffs, beverages and tobacco 
& 0.965 & 0.966 \\

V & Mineral products 
& 0.993 & 0.982 \\

VI & Products of the chemical or allied industries 
& 0.998 & 0.994 \\

VII & Plastics and rubber products 
& 1.003 & 0.997 \\

VIII & Raw hides, skins, leather and articles thereof 
& 1.026 & 1.028 \\

IX & Wood and articles of wood 
& 0.987 & 0.983 \\

X & Pulp, paper and printed products 
& 0.953 & 0.950 \\

XI & Textiles and textile articles 
& 0.992 & 0.998 \\

XIII & Articles of stone, cement, ceramics and glass 
& 0.994 & 0.990 \\

XV & Base metals and articles of base metal 
& 0.993 & 0.991 \\

XVI & Machinery, mechanical appliances and electrical equipment 
& 0.997 & 0.990 \\

XVII & Vehicles, aircraft and transport equipment 
& 0.972 & 0.974 \\

XVIII & Optical, medical and precision instruments 
& 1.022 & 1.012 \\

XX & Miscellaneous manufactured articles 
& 0.987 & 0.995 \\

\hline
Total &  & 0.990 & 0.988 \\
\hline
\end{tabular}

\vspace{0.2cm}

\begin{minipage}{\textwidth}
\footnotesize
\textit{Notes:} The table reports average CES bilateral tariff margins by HS section in 2016 and 2021. Values above one indicate bilateral tariffs lower than the CES reference tariff, while values below one indicate relatively weaker preferential access.
\end{minipage}

\end{table}

Table \ref{tab:country_btm} extends the analysis to country groups. Panel A reports average tariff margins by region. Most regions experience a decline between 2016 and 2021, consistent with a deterioration in relative market access. The largest reduction is observed in North America, where the average margin falls from 0.997 to 0.984. Declines are also recorded for Latin America and the Caribbean, South Asia, Europe and Central Asia, and the Middle East and North Africa.

By contrast, Sub-Saharan Africa is the only region exhibiting a slight improvement, with the average margin increasing from 1.003 to 1.004. In addition, it remains the only region with tariff margins consistently above unity in both years, indicating a relatively favourable position within the US tariff structure.

\begin{table}[H]
\centering
\caption{Average CES bilateral tariff margins across country groups}
\label{tab:country_btm}

\begin{tabular}{lcc}
\hline
\multicolumn{3}{c}{\textbf{Panel A. Regions}} \\
\hline
Region & 2016 & 2021 \\
\hline
East Asia \& Pacific & 0.986 & 0.985 \\
Europe \& Central Asia & 0.978 & 0.975 \\
Latin America \& Caribbean & 1.001 & 0.995 \\
Middle East \& North Africa & 0.989 & 0.987 \\
North America & 0.997 & 0.984 \\
South Asia & 0.997 & 0.992 \\
Sub-Saharan Africa & 1.003 & 1.004 \\
\hline
Total & 0.991 & 0.988 \\
\hline

\hline
\multicolumn{3}{c}{\textbf{Panel B. Income groups}} \\
\hline
Income group & 2016 & 2021 \\
\hline
High income: OECD & 0.986 & 0.986 \\
High income: non-OECD & 0.974 & 0.955 \\
Low income & 0.999 & 1.000 \\
Lower middle income & 0.996 & 0.998 \\
Upper middle income & 0.995 & 0.994 \\
\hline
Total & 0.991 & 0.988 \\
\hline
\end{tabular}

\footnotesize
\textit{Notes:} The table reports average CES bilateral tariff margins by region and income group in 2016 and 2021. Values above one indicate bilateral tariffs lower than the CES reference tariff, while values below one indicate relatively weaker preferential access.
\end{table}

Panel B reports the same indicators by income group. High-income non-OECD countries experience the strongest deterioration, with average margins declining from 0.974 to 0.955. Upper-middle-income countries record a small decline, while OECD economies remain broadly unchanged. By contrast, low-income and lower-middle-income countries display stable or slightly improving tariff margins, with average values close to or above unity.

These results suggest that changes in US tariff protection did not affect all exporters equally. While the overall decline in average tariff margins is relatively modest, the distribution of market access across sectors, regions, and income groups changed more substantially. This evidence is consistent with the idea that increases in MFN protection reshape the structure of tariff preferences and alter the relative competitiveness of exporters even before considering their effects on trade flows.

\subsection{Simulated trade effects}
Table \ref{tab:trump_scenario} summarizes the tariff measures included in the February 2026 simulation scenario. The scenario combines both economy-wide and sector-specific increases in US import protection. A general surcharge of 10 percent is imposed on imports from a broad group of trading partners, while Bolivia faces a higher economy-wide tariff increase of 15 percent. In addition, sector-specific measures target products that have been at the center of recent US trade policy debates, including steel and aluminum products (25 percent), automobiles and automobile parts (25 percent), and copper products (50 percent). The scenario therefore captures both the broad-based increase in protection and the concentration of tariff measures in selected manufacturing industries.

For each bilateral trade flow, the simulated tariff is obtained by adding the relevant surcharge to the observed tariff rate. The objective is not to predict the exact future evolution of trade flows, but rather to evaluate how higher MFN protection affects bilateral tariff margins, relative market access, and the allocation of imports across suppliers.

\begin{table}[H]
\centering
\caption{Trump tariff scenario (February 2026)}
\label{tab:trump_scenario}

\begin{tabular}{lc}
\hline
Measure & Additional tariff rate \\
\hline
Selected countries (economy-wide) & 10\% \\
Bolivia (economy-wide) & 15\% \\
Steel and aluminum products & 25\% \\
Copper products & 50\% \\
Automobiles & 25\% \\
Automobile parts & 25\% \\
\hline
\end{tabular}

\vspace{0.2cm}

\footnotesize
\textit{Notes:} The table summarizes the tariff increases included in the February 2026 simulation scenario. Economy-wide measures apply to all products imported from the affected countries, while sector-specific measures apply to all exporters. Source: Atlantic Council, \textit{Trump Tariff Tracker} (accessed May 2026).
\end{table}

Table \ref{tab:scenario1_results} reports the simulated trade effects by HS section. The results indicate that the increase in US tariffs generates both trade diversion and trade reduction, although the latter clearly dominates. Aggregate trade declines by approximately USD 93.3 billion, corresponding to a reduction of 7.9 percent relative to the baseline.

The largest losses are concentrated in a limited number of sectors. Optical, medical and precision instruments (XVIII) experience the strongest contraction, with imports falling by USD 48.1 billion (-38.0 percent). Chemicals (VI) record a reduction of USD 13.0 billion (-49.9 percent), while vehicles and transport equipment (XVII) decline by USD 23.5 billion (-22.9 percent). Significant negative effects are also observed for base metals (-11.4 percent), machinery and electrical equipment (-11.6 percent), and mineral products (-7.7 percent).

\begin{table}[H]
\centering
\scriptsize
\caption{Simulated trade effects of the February 2026 tariff scenario}
\label{tab:scenario1_results}

\begin{tabularx}{\textwidth}{lXccccc}
\toprule
HS & Sector &
Domestic effect &
3rd-country gain &
3rd-country loss &
Net effect &
Trade effect \\
&
&
(\%) &
(\%) &
(\%) &
(USD mn) &
(\%) \\
\midrule

IV &
Prepared foodstuffs, beverages and tobacco
& 0.0 & 4.2 & -0.3 & 2,835 & 3.9 \\

V &
Mineral products
& 0.0 & 1.4 & -9.1 & -5,504 & -7.7 \\

VI &
Products of the chemical or allied industries
& 0.0 & 0.0 & -49.9 & -12,995 & -49.9 \\

VII &
Plastics and rubber products
& 0.0 & 0.8 & -2.6 & -3,185 & -1.8 \\

VIII &
Raw hides, skins, leather and articles thereof
& -3.3 & 3.5 & -1.5 & -356 & -1.3 \\

IX &
Wood and articles of wood
& 5.6 & 0.0 & -1.8 & 1,036 & 3.8 \\

X &
Pulp, paper and printed products
& 0.0 & 0.0 & 0.0 & 0 & 0.0 \\

XI &
Textiles and textile articles
& -53.5 & 90.3 & -3.6 & 11,166 & 33.2 \\

XIII &
Articles of stone, cement, ceramics and glass
& -8.4 & 1.5 & 0.0 & -5,334 & -6.9 \\

XV &
Base metals and articles of base metal
& 0.0 & 0.0 & -11.4 & -14,041 & -11.4 \\

XVI &
Machinery, mechanical appliances and electrical equipment
& -16.7 & 5.2 & -0.1 & -6,766 & -11.6 \\

XVII &
Vehicles, aircraft and transport equipment
& 0.0 & 0.2 & -23.1 & -23,482 & -22.9 \\

XVIII &
Optical, medical and precision instruments
& -38.5 & 0.9 & -0.4 & -48,115 & -38.0 \\

XX &
Miscellaneous manufactured articles
& 9.5 & 0.1 & -1.2 & 11,404 & 8.5 \\

\midrule

Total &
All sectors
& -5.9 & 3.6 & -5.6 & -93,338 & -7.9 \\

\bottomrule
\end{tabularx}

\vspace{0.2cm}

\footnotesize
\textit{Notes:} The table reports simulated trade effects associated with the February 2026 tariff scenario. Net effects are expressed in millions of US dollars. Positive values indicate net trade expansion relative to the baseline, while negative values indicate net trade contraction. The final column reports the percentage change in trade relative to the baseline.

\end{table}
Not all sectors are negatively affected. Textiles and textile articles (XI) emerge as the main beneficiary of trade diversion. Imports from alternative suppliers increase by more than 90 percent, generating a net gain of approximately USD 11.2 billion and an overall increase of 33.2 percent relative to the baseline. Positive net effects are also observed for miscellaneous manufactured articles (USD 11.4 billion; +8.5 percent), prepared foodstuffs (USD 2.8 billion; +3.9 percent), and wood products (USD 1.0 billion; +3.8 percent).

The sectoral results suggest that higher tariffs induce substantial reallocation of imports across suppliers. However, in most industries the gains generated by trade diversion remain insufficient to compensate for the decline in imports from negatively affected exporters. Consequently, the overall effect of the tariff shock remains negative.

\begin{table}[H]
\centering
\small
\caption{Simulated trade effects by Broad Economic Categories (BEC)}
\label{tab:scenario1_bec}

\begin{tabular}{lccccc}
\toprule
BEC category &
Domestic effect &
3rd-country gain &
3rd-country loss &
Net effect &
Trade effect \\
&
(\%) &
(\%) &
(\%) &
(USD mn) &
(\%) \\
\midrule

Capital goods
& -9.5 & 0.9 & -3.4 & -29,872 & -12.1 \\

Consumption goods
& -2.0 & 7.9 & -7.6 & -6,706 & -1.7 \\

Intermediate goods
& -7.1 & 1.6 & -5.1 & -56,759 & -10.5 \\

\bottomrule
\end{tabular}

\vspace{0.2cm}

\footnotesize
\textit{Notes:} The table reports simulated trade effects of the February 2026 tariff scenario by Broad Economic Categories (BEC). Net effects are expressed in millions of US dollars. Positive values indicate net trade expansion relative to the baseline, while negative values indicate net trade contraction. The final column reports the percentage change in trade relative to the baseline.

\end{table}

Table \ref{tab:scenario1_bec} aggregates the results by Broad Economic Categories (BEC). The strongest effects are concentrated in intermediate and capital goods, highlighting the vulnerability of production networks to increases in trade protection.

Intermediate goods experience the largest contraction, with trade declining by USD 56.8 billion (-10.5 percent). Capital goods also record substantial losses, with imports falling by USD 29.9 billion (-12.1 percent). In contrast, consumption goods are comparatively less affected, showing a reduction of only USD 6.7 billion (-1.7 percent).

The decomposition of the trade effects provides additional insights. For consumption goods, imports from alternative suppliers increase by 7.9 percent, almost offsetting the losses experienced by affected exporters. By contrast, trade diversion is much weaker for intermediate and capital goods, where imports from alternative suppliers increase by only 1.6 and 0.9 percent, respectively. These results suggest that substitution across suppliers is considerably more difficult for production inputs and investment goods than for final consumer products.

The BEC analysis indicates that higher tariffs primarily disrupt supply chains and production-related trade flows rather than final consumption.

\begin{table}[H]
\centering
\small
\caption{Simulated trade effects by region and income group}
\label{tab:scenario1_regions_income}

\begin{tabular}{lcccc}
\toprule
Group &
Trade reallocation &
Trade diversion &
Net effect &
Trade effect \\
&
(\%) &
(\%) &
(USD mn) &
(\%) \\
\midrule
\multicolumn{5}{c}{\textbf{Panel A. Regions}} \\
\midrule

East Asia \& Pacific
& 17.7 & -15.2 & 2,907 & 2.5 \\

European Union
& 3.4 & -28.9 & -10,800 & -25.5 \\

Europe \& Central Asia
& 4.5 & -20.2 & -2,510 & -15.7 \\

Latin America \& Caribbean
& 18.6 & -22.2 & -2,061 & -3.6 \\

Middle East \& North Africa
& 13.5 & -13.7 & -30 & -0.2 \\

North America
& 3.0 & -19.5 & -9,972 & -16.5 \\

Others
& 6.1 & -25.9 & -3,274 & -19.8 \\

South Asia
& 33.8 & -14.7 & 2,235 & 19.1 \\

Sub-Saharan Africa
& 9.4 & -14.6 & -257 & -5.1 \\

United States
& -8.2 & 0.0 & -69,576 & -8.2 \\

\midrule

\multicolumn{5}{c}{\textbf{Panel B. Income groups}} \\
\midrule

High-income OECD economies
& 11.3 & -14.9 & -700 & -3.6 \\

High-income non-OECD economies
& 3.7 & -24.7 & -28,790 & -20.9 \\

Low-income economies
& 33.1 & -12.4 & 332 & 20.8 \\

Lower-middle-income economies
& 26.0 & -11.0 & 5,284 & 15.0 \\

Upper-middle-income economies
& 18.8 & -16.2 & 3,386 & 2.6 \\

\bottomrule
\end{tabular}

\vspace{0.2cm}

\footnotesize
\textit{Notes:} Trade reallocation combines the domestic adjustment effect and the increase in imports from alternative suppliers. Trade diversion measures the reduction in imports from negatively affected exporters. Net effects are expressed in millions of US dollars. The final column reports the percentage change in trade relative to the baseline.
\end{table}

Table \ref{tab:scenario1_regions_income} presents the geographical distribution of the simulated effects. The results reveal substantial heterogeneity across regions and income groups, generating both winners and losers from the tariff reform.

Among regions, the largest losses are concentrated in the European Union and North America, where trade declines by 25.5 and 16.5 percent, respectively. Europe and Central Asia also experience a sizeable contraction (-15.7 percent). These regions account for a large share of US imports and are therefore particularly exposed to the increase in tariff protection.

By contrast, South Asia emerges as the principal beneficiary of trade diversion, with trade increasing by 19.1 percent. East Asia and the Pacific also record a positive net effect (+2.5 percent), while the Middle East and North Africa remain broadly unaffected. The results indicate that part of the trade displaced by higher tariffs is redirected toward alternative suppliers located in developing regions.

The income-group decomposition reveals a similar pattern. High-income non-OECD economies experience the largest losses, with trade declining by USD 28.8 billion (-20.9 percent). In contrast, low-income and lower-middle-income economies benefit from trade diversion, recording increases of 20.8 and 15.0 percent, respectively. Upper-middle-income economies also experience a modest positive effect (+2.6 percent).

These findings provide strong evidence that increases in MFN protection reshape the distribution of market access across exporters. Trade diversion generates gains for some sectors, regions, and income groups, particularly among developing economies. Nevertheless, these gains are not sufficient to offset the losses experienced by the most exposed exporters. As a result, aggregate trade declines by approximately USD 93.3 billion, confirming that the contractionary effects of higher tariffs dominate the reallocation of trade toward alternative suppliers.

\section{Conclusion}

This paper examines the effects of increases in MFN tariffs on US imports through the lens of bilateral tariff margins and relative market access. Building on a structural gravity framework with CES-consistent tariff margins, we show that changes in MFN protection modify not only the level of trade costs but also the relative value of existing trade preferences. As MFN tariffs increase, exporters benefiting from preferential access experience an improvement in their relative competitiveness even though their preferential tariff treatment remains unchanged. We refer to this mechanism as \emph{preference amplification}.

The empirical analysis confirms that higher MFN protection generates both trade diversion and trade contraction. Estimated substitution elasticities display substantial heterogeneity across products, implying that the response to changes in tariff margins differs markedly across sectors. Consequently, the effects of protection are far from uniform. While some exporters benefit from improved relative market access, others experience significant losses, generating pronounced differences across industries, regions, and income groups.

The counterfactual analysis based on the February 2026 US tariff scenario illustrates the magnitude of these effects. Aggregate US imports decline by approximately USD 93 billion, corresponding to a reduction of almost 8 percent relative to the baseline. The largest losses are concentrated in intermediate and capital goods, particularly in chemicals, transport equipment, and optical and precision instruments, suggesting that higher tariffs disrupt production networks more than final consumption. By contrast, sectors such as textiles and miscellaneous manufactured articles experience relatively smaller losses, as part of the reduction in imports from non-preferred suppliers is offset by increased imports from preferential trading partners. Similar patterns emerge across regions and income groups, with South Asia and several developing economies expanding their market shares at the expense of exporters more heavily affected by the tariff increases.

These findings suggest that modern protectionism operates through two interconnected channels. On the one hand, higher MFN tariffs reduce aggregate trade by increasing trade costs and lowering imports from non-preferred exporters. On the other hand, they simultaneously redistribute market opportunities by strengthening the relative competitiveness of exporters benefiting from preferential access. The gains associated with this reallocation, however, are not sufficient to compensate for the overall contraction in trade. Trade diversion therefore mitigates, but does not offset, the contractionary effects of higher protection.

The analysis is subject to several limitations. First, the empirical framework is estimated from the perspective of a single importing country and therefore captures only partial-equilibrium adjustments, although the CES bilateral tariff margins explicitly account for the multilateral tariff structure faced by competing exporters. Second, the counterfactual simulations abstract from broader adjustments in prices, production, investment, exchange rates, and firms' location decisions. Consequently, the estimated effects should be interpreted as the trade responses generated by changes in relative market access rather than as precise forecasts of future trade outcomes.

These limitations also point to promising directions for future research. Extending the analysis to a multi-country structural gravity framework would make it possible to evaluate the interaction between MFN protection, preferential market access, and multilateral resistance in general equilibrium. More broadly, the concept of \emph{preference amplification} introduced in this paper may provide a useful framework for analysing a wider range of trade policy reforms in which changes in MFN tariffs modify the effective value of existing preferential agreements without altering the agreements themselves. As protectionism and preferential trade agreements increasingly coexist in the global trading system, accounting for this mechanism may improve our understanding of how trade policy reshapes market access, trade diversion, and the international allocation of production.

\bibliographystyle{apalike}
\bibliography{References.bib}
\end{document}